\documentclass[a4paper,aps,preprintnumbers,twocolumn,amsmath,amssymb,prl]{revtex4}
\usepackage{bbm}
\usepackage{mathrsfs}
\usepackage{amsmath}
\usepackage[dvips]{graphicx}
\usepackage{epsfig}
\usepackage[dvips]{graphics,graphicx}

\newcommand{\be}{\begin{equation}}
\newcommand{\ee}{\end{equation}}
\newcommand{\bea}{\begin{eqnarray}}
\newcommand{\eea}{\end{eqnarray}}
\newcommand{\bes}{\begin{split}}
\newcommand{\ees}{\end{split}}

\usepackage[dvips]{graphics,graphicx}

\begin{document}
\title{Dynamical Structure Factor and Spin-Density Separation for a Weakly-Interacting Two-Component Bose Gas}
\author{ M.-C. Chung and A. B. Bhattacherjee}
\affiliation{Max-Planck-Institut f\"ur Physik komplexer Systeme,
01187 Dresden, Germany}

\begin{abstract}
  We show that spin-density separation in a Bose gas is not restricted to 1D but also occurs in higher dimension. The ratio ($\alpha$) of the intra-species atom-atom interaction strength to the inter-species interaction strength, strongly influences the dynamics of spin-density separation and the elementary excitations. The density wave is phonon-like for all values of $\alpha$. For $\alpha<1$, spin wave is also phonon-like. The spin waves have a quadratic dispersion in the $\alpha=1$ coupling regime, while in the phase separated regime ($\alpha>1$) the spin waves are found to be damped. The dynamical structure factor (DSF) reveals two distinct peaks corresponding to the density and spin waves for $\alpha \le 1$. For $\alpha > 1$ there is only one DSF peak corresponding to the density wave. 
\end{abstract}

\date{\today}
\maketitle


Spin-density(charge) separation is a remarkable feature predicted for one-dimensional interacting spin-$1/2$ fermions \cite{Giamarchi} and widely searched in condensed matter systems. Consequently, its investigation in atomic systems could be of interest to different fields of physics. Unlike the higher-dimensional fermionic systems,  where elementary excitations normally carries both spin and density(charge) degree of freedom, the collective excitations of the one-dimensional Fermi system separate into two distinct modes, spin and density waves due to the fact that the interaction in one-dimensional system lead to a Luttinger liquid state with bosonic excitations \cite{Giamarchi}.  This behaviour is a hallmark of collective effects caused by interactions. For bosons, using two-components, corresponding to two hyperfine states of cold atoms \cite{Hall} allows us to study the (iso)spin waves as the relative spatial oscillations of the two-components. Till now, the study of spin-density separation has been limited to one-dimensional fermions \cite{Recati} and bosons \cite{Kleine, Fuchs} since, both these systems are believed to belong to the same Luttinger liquid class, which leads to the spin-density separation. Contrary to expectations, in this Letter we will show that for bosons spin-density separation not only exists in one-dimensional systems but also in higher dimensions since for bosons, excitations are always collective excitations in all dimensions due to the presence of the (quasi)condensate fraction and that the Luttinger liquid approach is not essential to describe the spin-density separation in bosons.  We also compute the dynamical structure factor which reveals distinct features of spin density separation in all dimensions.

We start with the Lagrangian density for
two-component Bose gas at zero temperature:
 \be \label{LagrangianO}
  \begin{split}
  \mathcal{L} =  & \frac{i}{2} \sum_{i=1,2} (\varphi_{i}^{\star} \partial_t
  \varphi_i - \varphi_{i} \partial_t \varphi_i^{\star}) -
  \frac{1}{2m} (\nabla_r\varphi_i)^2 - \mu_i n_i \\
    - & \frac{1}{2} \sum_{i,j=1,2} g_{d,ij} n_i n_j,
   \end{split}
  \ee
 where $\varphi_i = \varphi_i(r,t), i=1,2$ is the field representing two different Bose particles,
 $r$ is the space coordinate, $t$ is the real time, and here we set $\hbar=1$. Also $\mu_i$ , $n_i=|\varphi_i|^2$ is the chemical potential and the particle density of the $i^{th}$ component and  $g_{d,ij} > 0$ is the repulsive effective atom-atom
 interaction between the $i^{th}$ and $j^{th}$ components. Here we consider bosonic atoms of the same isotope of mass $m$ but having different internal spin states, therefore we have
 $g_{d,11}=g_{d.22}\equiv g_d$ and $g_{d,12}=g_{d,21} \equiv g_{d}'$.
 For simplicity, we consider the same average
 atom number density for the two components, i.e. $\bar{n}_1 = \bar{n}_2 = \bar{n}$.
  For a 3D Bose gas , $g_3 = 4\pi a_3/m$ \cite{LeeHuangYang,Leggett}. Here $a_{3}$ is the 3D scattering length.
  For lower dimensional Bose gas in a 3D trap with
  longitudinal harmonic trapping frequency $\omega_{\perp}$, $g_2 = 4\pi/(m \ln \bar{n}
  a_2)$ \cite{Popov,TwoDInt, Fisher} with the 2D scattering length given as $a_2 \simeq 7.41 e^{-\sqrt{\pi/a_3^2 m
  \omega_{\perp}}}$ \cite{Lee} and  $g_1 = 2 \omega_{\perp} a_{3}$ with $a_3 \ll  1/\sqrt{m
 \omega_{\perp}}$ \cite{Olshanii}.
  The behavior of the system depends
 crucially on the dimensionless parameter $\gamma_d (\gamma'_d) = m g_d n^{1-2/d} (m g'_d n^{1-2/d}
 )$. For the gas to be weakly interacting, we must have $\gamma_d(\gamma'_d) \ll 1$.
 The chemical potential $\mu_i$ of the $i^{th}$ component is determined by the condition $\sum_{ij} g_{d,ij} \bar{n}_j = \mu_i \bar{n}_i$.

To understand the low-energy excitations in two-component Bose gas,
one can derive a low-energy effective hydrodynamical Lagrangian that
contains only modes related to the low-energy excitations
\cite{Popov, Wen}. We write the Boson field $\varphi_i$ in the terms of the number density $n_{i}$ and the phase $\theta_i$ as $\varphi_i = n_i e^{i\theta_i}$. 
In the weak coupling regime the phase changes slowly in space
while the density fluctuates fast\cite{Leggett}, therefore one can
integrate out the high energy fast density fluctuation \cite{Wen} to
obtain the effective hydrodynamic action. We introduce the density fluctuation $\delta
n_i$ as $n_i = \bar{n} + \delta n_i$. In terms of the new basis, $\delta
n_{\rho(\sigma)} = (\delta n_1 \pm \delta n_2)/\sqrt{2}$ and
$\theta_{\rho(\sigma)} = (\theta_1 \pm \theta_2)/\sqrt{2}$, the action obtained from the  Lagrangian density
can be rewritten as
\be \label{action}
 \begin{split}
  S = & -\int \mbox{d}^d\mathbf{x} \mbox{d}t \sum_{\lambda = \rho,
 \sigma} [(\bar{n}_{\lambda} +\delta n_{\lambda})
 \partial_t\theta_{\lambda}  \\
  + &\frac{\bar{n}(\nabla_r\theta_{\lambda})^2}{2m}  + \frac{(\nabla_r
  \delta n_{\lambda})^2}{8m\bar{n}} + \frac{g_{d,\lambda}}{2} (\delta
  n_{\lambda})^2 ] ,
  \end{split}
\ee where $\bar{n}_{\rho} (\bar{n}_{\sigma}) = \sqrt{2} \bar{n} (0)$ and
$g_{d, \rho(\sigma)} = g (1 \pm \alpha)$ with $\alpha =
g'_{d}/g_{d}$. 

\begin{figure}
\center
\includegraphics[width=7.0cm]{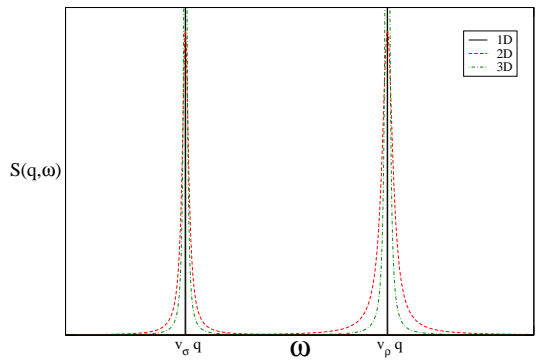}
\caption{Dynamic structure factor for all dimensions in the hydrodynamic regime and at zero temperature for $\alpha =0.5 $ and $q=1.25$, shows two distinct peaks corresponding to the density and the spin  waves, centered at $v_{\rho} q$ and $v_{\sigma} q $, respectively. The one dimensional structure factor is found to be a delta  function, while the two- and three dimensional DSF is broadened because of Beliaev damping. In two dimension the DSF for the density wave is broader compared to that of spin waves, while the width of the three-dimensional peaks remains the same. There is no specific reason why the width of the $2D$ peaks is  larger than both the $1D$ and $3D$ peaks, this depending on the choice of the parameters .} \label{fig1}
\end{figure}

For $\alpha < 1$, after performing two Gaussian integrals, the
effective action has the form\cite{Wen}
 \be\label{twosoundsaction}
 S_{\mbox{eff}} = \int \mbox{d}^d\mathbf{x} \mbox{d}t
 \sum_{\lambda = \rho,\sigma}
   \frac{\chi_{\lambda}}{2} \left(|\partial_t\phi_{\lambda}|^2 -
   v_{\lambda}^2 |\nabla_r\phi_{\lambda}|^2\right), \ee
 where $\phi_{\rho(\sigma)} = e^{i\theta_{\rho(\sigma)}}$,  $\chi_{\rho(\sigma)} (= 1/g_{d,\rho(\sigma)})$ is the density (spin) compressibility and
 $v_{\rho(\sigma)} (= \sqrt{\bar{n} g_{d,\rho(\sigma)}/m}) $ is the sound velocity of the density(spin) mode. Here we assumed that the fields $\theta_{\rho(\sigma)}$ vary slowly in space
 and we have dropped the $\frac{\nabla_r^2}{8m\bar{n}}$ term.
 The effective action (\ref{twosoundsaction})
 describes the low-energy excitations of two sound waves with
 linear dispersions $\omega_{\rho(\sigma)} = v_{\rho(\sigma)}  k $. The
 bosons split into two gapless modes, namely density mode and spin mode,
 propagating with different velocities. The density wave propagates
 faster than the spin wave, which can be seen by the relation $v_{\sigma}/v_{\rho} \approx
 \sqrt{(1-\alpha)/(1+\alpha)}$. 
 In this regime  the energy gap of the lowest excitation above the ground state is zero. Such systems have a diverging length scale determing the exponential decay of equal time correlatons in the ground state, which defines the quantum critical behavior. Therefore the systems for $\alpha < 1$ lie at the quantum critical points \cite{Sachdev}. 

 The meaning of the low-energy effective Lagrangian
 (\ref{twosoundsaction}) is that the bosonic system separates into two
 independent degree of freedom, i.e. spin and density. Unlike in fermionic one-dimensional systems, we do not need the bosonization method to obtain the spin-density separation, the only thing we need is the (quasi)condensate density $\bar{n}$ to have fluctuations around it.  
This can be fulfilled in all dimensions at zero temperature for bosonic systems.  The Bogoliubov energy dispersion
 relation of one-component interacting Bose gas is $\epsilon(k) = \sqrt{((k^2/2m)^2 + g_{d}\bar{n}k^2/m )}$
 \cite{Bogoliubov}. For the two component Bose gas, replacing the interaction $g_{d}$ with
 $g_{d,\rho(\sigma)}$, we obtain two branches of the excitations
  \be \label{dispersionep}
  \epsilon_{\rho(\sigma)}(k) = \sqrt{(k^2/2m)^2 + \frac{g_{d,\rho(\sigma)}(1 \pm \alpha)
  \bar{n}}{m}k^2},
 \ee
  which is in agreement with the result obtained by the
  semiclassical method \cite{Alexandrov}. From the dispersion
  relations (\ref{dispersionep}) we can define the chemical
  potential for the density and spin waves as $\mu_{\rho(\sigma)} = g_{d,\rho(\sigma)} \bar{n}$.

  For $\alpha =1 (g_{d} = g'_{d})$, only one Gaussian integral can be performed in
  action (\ref{action}) giving the gapless density wave with linear
  dispersion. However, one obtains a quadratic dispersion for the
  spin-wave excitations, in agreement with $SU(2)$ symmetry \cite{Halperin, Fuchs}. This effect can also
  be seen from the Bogoliubov excitations $\epsilon_{\rho} = \sqrt{(k^2/2m)^2 + 2g\bar{n}k^2/m}$
  and $\epsilon_{\sigma} = k^2/2m$ by replacing $g_{d,\rho} = 2g_{d}$ and
  $g_{d,\sigma} =0$ in (\ref{dispersionep}).
  In this case, due to the $SU(2)$ symmetry, the eigenstates are classified according to
  their total spin $S$ ranging from 0 to $N/2$,  and according to recent result by Eisenberg and Lieb \cite{Eisenberg},
  the ground state is fully polarized ($S=N/2$). In one dimension, the ground state is
  described by Lieb-Liniger(LL) model of one-component interacting Bose gas
  \cite{LiebLiniger}, for which the elementary excitations in the
  weak-coupling regime are density waves
  \cite{Lieb}, and the system is ferromagnetic.
  
  In the case of $\alpha >1 (g_{d}<g'_{d})$, we found $g_{d,\sigma} <0$. This
  implies, $v_{\sigma}$ ($= \sqrt{g_{d,\sigma}\bar{n}/m}$) in the long wave length limit is imaginary.
  The spin waves become unstable and damp out in the thermodynamic
  limit. Therefore we obtain a phase separation of the two-component
  Bose gas \cite{Alexandrov}.

  The dynamical structure factor (DSF) of many-body system is defined as
  follows
  \be \label{DSF}
     S_{\rho(\sigma)}(q, \omega) = \int \mbox{d}^d\mathbf{x}
     \mbox{d}t e^{i(\omega t -{\mathbf q} \cdot {\mathbf x} )}
     \langle \delta n_{\rho(\sigma)}({\mathbf x},t) \delta n_{\rho(\sigma)}({\mathbf
     0},0) \rangle,
  \ee
 where $\langle \cdots \rangle$ can be calculated using path
 integral with the effective action.
  Experimentally, one can measure the
  dynamical structure factor using Bragg spectroscopy\cite{Steinhauer}. 

  For $\alpha < 1$, from the action \ref{action}, one can
  get the equation of motion for $\delta n_{\rho(\sigma)}$ as
  $\delta n_{\rho(\sigma)}({\mathbf x},t) = -1/g_{d,\rho(\sigma)}
  \partial_t
  \theta_{\rho(\sigma)}({\mathbf x},t)$. From the quadratic Lagrangian density,
  the DSF (\ref{DSF}) can be obtained as
    \be \label{DSFTwosound}
    S_{\rho(\sigma)}(q, \omega)  =
  \mbox{Im} \frac{\chi_{\rho(\sigma)} v^2_{\rho(\sigma)} q^2}{\omega^2 - \omega^2_{\rho(\sigma)}(q)}, \ee
  where $\omega_{\rho(\sigma)}(q) = v_{\rho(\sigma)} q +i \Gamma_{\rho(\sigma)}(q)$ with the quasiparticle  decay rate
  $\Gamma_{\rho(\sigma)}(q)$.

 In order to obtain the DSF, one has to find the compressibility
 $\chi_{\rho(\sigma)}$, velocity $ v_{\rho(\sigma)}$ and decay rate $\Gamma_{\rho(\sigma)}(q)$ in terms of the dimensionless parameters $\gamma_{d,
 \rho(\sigma)}$. Using the macroscopic argument, the compressibility
 $\chi_{\rho(\sigma)}$ is related to the energy $E(g_{\rho(\sigma)}, n)$ as $\chi_{\rho(\sigma)}^{-1} =
 \frac{1}{V}  \frac{\partial^2E}{\partial
 n^2}$ with the constant system size: $V=L^d$ and density: $n = N/V$.
 Similarly, the sound velocity can also be obtained using the
 macroscopic energy spectrum as $v_{\rho(\sigma)} = (\frac{V}{mn} \frac{\partial^2E}{\partial
 V^2})^{1/2}$ with constant particle number $N$ \cite{London}. The
 way to obtain the ground state energy spectrum is diverse and
 depends on the dimension. 
 As indicated by Beliaev \cite{Beliaev}, the dimensional dependent decay rate $\Gamma_{\rho(\sigma)}(q)$
 is caused by the process of a long wave-length phonon decaying into
 two phonons and it can be calculated for small momenta using the formula \cite{Popov}
 \be {\label{Beliaevdamping}}
   \begin{split}
   \Gamma_{\rho(\sigma)}(q) = & \frac{9
   v_{\rho(\sigma)}}{128\pi^2\bar{n}m} \int {\mbox d}^d k |{\mathbf q}| |{\mathbf
   k}| |{\mathbf q}-{\mathbf k}| \\ & \delta(\epsilon_{\rho(\sigma)}({\mathbf
   q}) - \epsilon_{\rho(\sigma)}({\mathbf  k}) - \epsilon_{\rho(\sigma)}({\mathbf
   q}-{\mathbf k})).
   \end{split}
 \ee
 
 In 3D, the dimensionless parameter $\gamma_{3} = 4\pi a_{3} \bar{n}^{1/3}$. In this case, the requirement
 for  a dilute gas  $\bar{n} a^3 \ll 1$ corresponds
 to the weak-coupling condition $\gamma_{3} \ll 1$. The ground state
 energy was given for the first time by Lee et al. \cite{LeeHuangYang} as
 $ E = N \bar{n}^{2/3}/(2m)  \gamma_3 (1+16 \gamma_3^{3/2}/5\pi^2).$
 The ground state compressibility and velocity are given by $\chi_{\rho(\sigma)}^{-1}  =  g_{3,\rho(\sigma)} (1+\frac{2}{\pi^2}
   \gamma_{3,\rho(\sigma)}^{\frac{3}{2}})$ and $
    v_{\rho(\sigma)}  =   \sqrt{\frac{g_{3,\rho(\sigma)}
    \bar{n}}{m}}
    (1+\frac{2}{\pi^2}
   \gamma_{3,\rho(\sigma)}^{\frac{3}{2}})^{\frac{1}{2}},$
   respectively.
 The decay rate for 3D system is obtained from
 eq(\ref{Beliaevdamping}): $\Gamma_{\rho(\sigma)}(q) = \Gamma(q) = \frac{3 q^5}{640 \pi m
  \bar{n}}$ \cite{Beliaev}.
 We can see that the decay rates for density and spin waves are equal and proportional to
 $q^5$. The DSF for $\omega > 0$ can be
 approximated as
 \be \label{ImDSF}
    S_{\rho(\sigma)}(q, \omega)  \approx
   \frac{\chi_{\rho(\sigma)} v_{\rho(\sigma)} q \Gamma_{\rho(\sigma)}(q)}{2\left[(\omega - v_{\rho(\sigma)}
   q)^2+\Gamma_{\rho(\sigma)}^2(q)\right]},
 \ee
 In the Bragg scattering experiment, one should obtain two peaks centered at $v_{\rho(\sigma)} q$ for the cross
 section with the width $\Gamma(q)$.

 For 2D Bose gas,renormalization-group analysis
 \cite{Fisher,Kolomeisky} shows that the interaction of the $2D$ dilute gas
 is marginally irrelevant only in a dilute limit specified by $\ln{\ln{\gamma_{2}}} \gg
 1$. The corresponding  ground state energy for a weak-interacting gas is
given by  $E = N \bar{n} /(2m) \gamma_{2} (1- C \gamma_{2})$
  where constant $C\ll 1$ is not universal but model-dependent due to the marginal interaction\cite{Kolomeisky}.
  The compressibility and velocity
for spin and
 density-wave excitations are $\chi^{-1}_{\rho(\sigma)} = g_{2,\rho(\sigma)}(1-(C-\frac{3}{8\pi})\gamma_{2,\rho(\sigma)})$
  and $v_{\rho(\sigma)} = \sqrt{\frac{\bar{n}\gamma_{2,\rho(\sigma)}}{m}}
  (1-(C-\frac{3}{8\pi})\gamma_{2,\rho(\sigma)})^{1/2}$.
  The Belieav decay rate can be obtained by the integral
  (\ref{Beliaevdamping}): $\Gamma_{\rho(\sigma)}(q) = \frac{\sqrt{3}
  v_{\rho(\sigma)} q^3}{64\pi \bar{n}}$.  Therefore the  DSF
  (\ref{ImDSF}) has a broader width for density waves than spin waves.

 In the case of one dimension, contrary to 2D and 3D systems,
  the weak coupling means that the system is in the high density regime
 because $\gamma_{1} = mg_{1}/\bar{n}$. In this regime,
  Lieb and Liniger \cite{LiebLiniger} first gave the
  ground state energy as
   : $E = \frac{N n^2}{2m} \gamma_{1}(1-\frac{4}{3\pi}
   \sqrt{\gamma_{1}})$. A few algebra leads to the compressibility
   and sound velocities as $\chi^{-1}_{\rho(\sigma)} = g_{1,\rho(\sigma)}(1-\frac{1}{2\pi}
   \sqrt{\gamma_{1,\rho(\sigma)}})$ and $v_{\rho(\sigma)} = \sqrt{\frac{g_{1,\rho(\sigma)} \bar{n}}{m}} (1-\frac{1}{2\pi}
   \sqrt{\gamma_{1,\rho(\sigma)}})^{1/2}$. For 1D, one obtains  no decay
   rate. The reason is  that the scenario for one phonon decaying into
   two phonons cannot exist due to the fact that energy conservation law in
   eq(\ref{Beliaevdamping}) cannot be fulfilled in 1D. Therefore two
   sharp peaks should be observed in the Bragg scattering experiments. Figure \ref{fig1} illustrates and summarizes the results obtained above for the DSF in all the three dimensions. 

\begin{figure}
\center
\includegraphics[width=10cm]{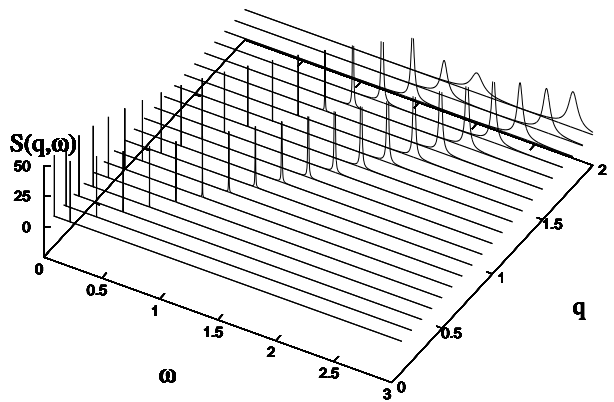}
\caption{Dynamic structure factor for $SU(2)$ in symmetric Hamiltonian in $3D$.  Note that $\omega$ is in the units of $g_d \bar{n}$ and $q$  is in the units of $\sqrt{m g_d \bar{n}}$.  The DSF of the density waves varies linearly with $q$, while the DSF of the spin waves shows the quadratic  dependence on $q$. } \label{fig2}
\end{figure}

  In the case of $\alpha = 1$, the situation changes. For  the density waves
the dynamic structure factor remains the same as that in two-sound regime,
while the DSF for spin-wave exitation  alters due to the dramatic changing of
the dispersion from linear to quadratic. In order to calculate the DSF one can
use the effective Hamiltonian in the weak-coupling regime: 
\be \label{HSU2}
  \begin{split}
  H = & \sum_p \epsilon_p a_p^{\dagger} a_p + \sum_p e_p b_p^{\dagger}
  b_p  \\    + & g_d \sqrt{\frac{\bar{n}}{V}} \sum_{k,q \neq 0} \sqrt{\frac{e_q}{\epsilon_q}}(a_q^{\dagger}+a_q)
  b_{k-q}^{\dagger} b_k 
  \end{split}
\ee
 with the spectrum of free spin waves $e_p=p^2/2m$, the Bogoliubov
spectrum $\epsilon_p = \sqrt{e_p^2+2\mu_d n e_p}$ \cite{Fuchs} and the chemical potential : $\mu_d = 2 g_d \bar{n}$ .   
Using $\delta n_{\sigma} = \sqrt{\bar{n}} (b^{\dagger}+b)$,
the DSF can be related to the imaginary part of the Green function as
$S(q,\omega) = \bar{n} \mbox{Im} G(q,\omega)$  where $G(q,\omega)$ is the singe particle Green function of the spin operators $b_q$ and $b^{\dagger}_q$.
Therefore the DSF for the spin waves reads 
\be \label{DSFSU2}
    S(q,\omega) = \frac{\bar{n} \Gamma_{d,\sigma}(q)}{(\omega-\frac{q^2}{2 m_d^{\star}})^2 +  \Gamma_{d,\sigma}^2(q) } , 
\ee
where the effective mass $m^{\star}$ is determined by the equation: $m/m_d^{\star} = (1+ 2/m \partial ^2 \mbox{Re} \Sigma(p)/\partial p^2) (p=0)$ with the self energy defined as $\Sigma = G^{-1} (g_d)- G^{-1}(g_d=0)$,  and the decay rate:  $ \Gamma_{d,\sigma}(q)  = \mbox{Im} \Sigma(q)$.   To the second order diagram for the self energy $\Sigma$, one obtains the inverse effective mass  related to the dimensionless parameter  $\gamma_d = \mu_d \bar{n}^{-2/d} $ as  $m/m_d^{\star} = 1-\alpha_d \gamma_d^{d/2}$ with $\alpha_d = 2/3\pi, 1/2\pi, 1/8\pi$ for one, two and  three dimensions, respectively.   
 The decay process  depends on the spin-phonon interaction which requires the
 energy conservation: $e_{q-k}+\epsilon_k = e_q$ with the spin momentum $q$
 and phonon momentum $k$. For $q< \sqrt{m \mu_d}$, this condition cannot be fulfilled, therefor $\Gamma_{d \sigma} = 0$, i.e. 
$S(q,\omega) = \bar{n} \delta(\omega - \omega_q)$. For $q \gtrsim \sqrt{m \mu_d}$ , 
an aproximation  can be obtained as follows : $\Gamma_{d\sigma}(\sqrt{m\mu_d} (1+\delta)) = \beta_d \mu_d \delta^{3/2 (d-1)} $ for $\delta \ll 1$ with $\beta_d = 0, 1/8\pi, 2/3\pi$ for $d=1,2,3$, respectively.    
The eqn. (\ref{DSFSU2}) shows the fact that the excitations for spin waves
 for the Hamitonian with  $SU(2)$ symmetry are not sound-like, but
 particle-like, with the DSF centered at the position proportional to $q^2$
 instead of $q$. Similar to the two-sound mode regime ($\alpha <1$), the DSF
 for one dimension is a delta function due to the fact that the energy
 conservation relation for a particle emitting a phonon cannot be fulfil in one
 dimension at zero temperature.  Fig. \ref{fig2} shows the linear dispersion of the DSF for  the density waves and the quadratic dispersion of the DSF for the spin waves. The delta function behavior are shown for low momenta.

For phase-separation regime ($\alpha >1$), the spin waves are thermodynamicall unstable, therefore only density waves exists. The dynamic structure factor of spin waves is smeared out and there exists only one peak in the DSF, which is different from the other regimes. This property can be a prominent signature for checking whether the system is phase separated or not.


In this Letter we have shown that, unlike fermionic systems, spin-density separation in two-component bosonic system is a more generic feature and occurs in all dimensions.  The density wave is found to be phonon like for all dimensions and coupling regimes. However, the spin waves in the two-sound regime ($\alpha<1$) show a linear dispersion (phonon like) and the DSF for all dimensions show two distinct peaks corresponding to the density and the spin  waves, centered at $v_{\rho} q$ and $v_{\sigma} q $, respectively. In the same regime,  the one-dimensional structure factors are  found to be  delta  functions, while the two- and three-dimensional DSF is broadened because of Beliaev damping.  The spin waves show a quadratic dispersion in the SU(2) symmetric regime ($\alpha=1$) and the DSF for all dimensions also show two distinct peaks centered at $v_{\rho} q$ and $q^2/2m_{d}^{\star}$. The spin wave is damped in the phase separated regime ($\alpha>1$) and there is only one peak corresponding to the density wave.  These are interesting signatures of spin-density separation to look for using Bragg spectroscopy where, the response of the condensate to a two-photon Bragg pulse is measured\cite{Steinhauer}. The difference between spin and charge velocities allows us to have spin and charge wavepackets moving at different velocities. An optical potential generated by a laser tuned,e.g., between fine-structure levels of excited alkali states transfers momentum solely to the the spin waves, while an optical potential far detuned will act solely on the density waves \cite{Recati}. One can also coherently excite the spin waves and the density waves simultaneously and then probe the two waves with a second laser pulse at a later time. Spin-charge separation manifests itself in a spatial separation of the spin and density wavepackets.

\bigskip


\end{document}